\documentclass[pdflatex,sn-nature]{sn-jnl}

\usepackage{hyperref}
\usepackage{tabularx}
\usepackage{graphicx}
\usepackage{multirow}%
\usepackage{amsmath,amssymb,amsfonts}%
\usepackage{amsthm}%
\usepackage{mathrsfs}%
\usepackage[title]{appendix}%
\usepackage{xcolor}%
\usepackage{textcomp}%
\usepackage{manyfoot}%
\usepackage{booktabs}%
\usepackage{algorithm}%
\usepackage{algorithmicx}%
\usepackage{algpseudocode}%
\usepackage{listings}%

\graphicspath{{./figures/}}
\newcommand{\blackboxname}{property net}
\newcommand{\tabref}[1]{table \ref{#1}}
\newcommand{\figref}[1]{figure \ref{#1}}
\newcommand{\secref}[1]{section \ref{#1}}

\newcommand{\AUCPO}{$\rm{AUC_{PO}}\ $}
\newcommand{\AUCIV}{$\rm{AUC_{IV}}\ $}
\newcommand{\Cmax}{$\rm{C_{max,PO}}\ $}
\newcommand{\AUCday}{$\rm{AUC_{24h}}\ $}
\begin{document}
\title{A Deep Neural Network -- Mechanistic Hybrid Model to Predict Pharmacokinetics in Rat}

%% or include affiliations in footnotes:
\author*[1]{Florian F{\"u}hrer}\email{florian.fuehrer@bayer.com}
\author[2]{Andrea Gruber}\email{andrea.gruber@bayer.com}
\author[3]{Holger Diedam}\email{holger.diedam@bayer.com}
\author[4]{Andreas H. G{\"o}ller}\email{andreas.goeller@bayer.com}
\author[2]{Stephan Menz}\email{stephan.menz@bayer.com}
\author[1]{Sebastian Schneckener}\email{sebastian.schneckener@bayer.com}
\affil[1]{\orgdiv{Engineering \& Technology, Applied Mathematics},
            \orgname{Bayer AG},
            \orgaddress{\postcode{51368},
            \city{Leverkusen},
            \state{Germany}}}
\affil[2]{\orgdiv{Pharmaceuticals, R\&D, Preclinical Modeling \& Simulation},
            \orgname{Bayer AG},
            \orgaddress{\postcode{13353},
            \city{Berlin},
            \state{Germany}}}
\affil[3]{\orgdiv{Crop Science, Product Supply, SC Simulation \& Analysis},
            \orgname{Bayer AG},
            \orgaddress{\postcode{40789},
            \city{Monheim},
            \state{Germany}}}
\affil[4]{\orgdiv{Pharmaceuticals, R\&D, Molecular Design},
            \orgname{Bayer AG},
            \orgaddress{\postcode{42096},
            \city{Wuppertal},
            \state{Germany}}}

\abstract{
An important aspect in the development of small molecules as drugs or agrochemicals is their systemic availability after intravenous and oral administration. 
The prediction of the systemic availability from the chemical structure of a potential candidate is highly desirable, as it allows to focus the drug or agrochemical development on compounds with a favorable kinetic profile. However, such predictions are challenging as the availability is the result of the complex interplay between molecular properties, biology and physiology and training data is rare. 

In this work we improve the hybrid model developed earlier \cite{Schneckener2019}. We reduce the median fold change error for the total oral exposure from $2.85$ to $2.35$ and for intravenous administration from $1.95$ to $1.62$. This is achieved by training on a larger data set, improving the neural network architecture as well as the parametrization of mechanistic model.
Further, we extend our approach to predict additional endpoints and to handle different covariates, like sex and dosage form.
In contrast to a pure machine learning model, our model is able to predict new end points on which it has not been trained. We demonstrate this feature by predicting the exposure over the first 24h, while the model has only been trained on the total exposure.}
\keywords{
Hybrid modelling, Deep Learning, Property prediction, PBPK modelling, Drug design, Bioavailability, Pharmacokinetics}

\maketitle

\section{Introduction}\label{sec:introduction}
Drug discovery is about the optimization of the interaction of molecules with biological targets to achieve the desired therapeutic effect, while reducing toxical effects.
The same is true for developing compounds for applications in agriculture, with a large focus on the reduction of toxical effects in mammals.
Both development processes are long and risky. Numerous methods and tools have therefore been established to support decisions in the search for best performing candidates. 
Experimental characterization of compounds can take up considerable resources and time.
The targeted, early identification of favorable properties and consequently informed selection of compounds can significantly reduce development cycles and the associated costs. 
Selection criteria include both, pharmacological and toxicological effects, as well as pharmacokinetics (PK)\footnote{Even though the name Pharmacokinetics implies that the field is only concerned with pharmaceutical substances, the field is concerned with all types of xenobiotic substances, see \url{https://en.wikipedia.org/wiki/Pharmacokinetics}}, in particular the availability of the compound in the body.

In this multi-parameter optimization of  physicochemical properties, efficacy, safety and PK, many compounds are usually tested in different high-throughput assays to generate a basic understanding of a compound's characteristics. However, as PK is determined by the complex non-linear interplay of compound properties and physiology, using these assays to test and optimize all aspects and parameters relevant for PK is usually not possible. Therefore, animal studies remain an important contribution to understanding the PK characteristics of a potential drug candidate. However, animal studies are usually performed later in research for selected compounds that are already optimized with respect to the early accessible assays. This approach helps to keep the number of animal experiments low, but unfortunately often struggles from eventual limitations in further PK optimization.

The most import quantity in PK is the blood plasma concentration $C$ as a function of time after a an oral (per os, PO)  or intravenous (IV) administration. In this publication we are mainly interested in a few key parameters characterizing the concentration-time curve, such as the maximal concentration ($C_{max}$) and the exposure between two time points $t_1$ and $t_2$:
\begin{equation}
    AUC_{t_1,t_2} = \int_{t_1}^{_{t_2}}dt \: C(t). \label{eq:auc}\\
\end{equation}
Most important is the total exposure, i.e. the exposure between the time of administration and infinity, here simply denoted as $AUC$, sometimes also the exposure during the first 24h after administration \AUCday is considered. 
For pharmaceutical compounds, oral drug delivery plays an important role.
It represents the most common administration route and is convenient for patients and physicians leading to high patient compliance. The extent to which the systemic exposure of a drug after PO administration ($\rm{AUC_{PO}}$) differs from the exposure after intravenous (IV) administration ($\rm{AUC_{IV}}$) is quantified by the oral bioavailability $F$ defined as
\begin{equation} \label{eq:F}
    F = \frac{AUC_{PO}}{AUC_{IV}} \cdot \frac{D_{IV}}{D_{PO}},
\end{equation}
where $D_{PO}$ denotes the oral dose and $D_{IV}$ the IV dose. We would like to stress that \AUCPO, \AUCIV and hence $F$ depends on the compound as well as on the dose and the formulation. In general the $AUC$ is a non-linear function of the dose, depending on e.g. the metabolic capacity or the availability of binding proteins. In addition, the \AUCPO can have non-linear dose dependencies related to the oral absorption, e.g. due to a limited solubility or a transport mechanism. Typically, non-linearities in the dose dependence of \AUCIV can be neglected, this allows to extrapolate the \AUCIV from $D_{IV}$ to $D_{PO}$ hence it is sufficient to consider $D_{PO} = D_{IV}$.

Furthermore, the oral dosage form can affect the PK, i.e. whether the compound is administered as solution, suspension or tablet. The \AUCPO is typically lower for administration of a suspension or tablet than for a solution, since for a suspension or a tablet the particles first have to be released from the formulation and dissolve in the gastrointestinal tract (GIT). Note, that for tablets special so-called enabling formulations exist, which can increase the dissolution rate hence increase \AUCPO and $F$. Those formulations are usually not used in the early phases of drug and agrochemical development, hence we do not consider them in this publication.

While in pharmaceutical development a compound's systemic availability is desired to be as high as possible, for agrochemicals a compounds' it should be as low as possible to minimize the risks associated with safety and toxicities.

As the determination of the PK-parameters $AUC$, $C_{max}$ and $F$ requires performing in-vivo studies in animals or even humans, they can not be used as a selection criterion for the early screening and optimization phases due to effort, cost and animal welfare considerations. Therefore, being able to predict them as early as possible, preferably directly from a compound's chemical structure, would reduce the risk during lead identification and optimization phases. Equally important, focusing only on the most promising compounds reduces the number of animal experiments.

There have been several attempts to predict PK-parameters from chemical structure \cite{Tian2011,FalconCano2020}. However, most of them are purely data-driven, hence do not exploit the available mechanistic knowledge about the different processes determining PK. In the present work, we combine Deep Learning with a mechanistic model to predict $AUC_{PO}$, \AUCIV and \Cmax in rats from the chemical structures only. Predictions for $F$ can be calculated using \eqref{eq:F} from the prediction for \AUCPO and \AUCIV. 

Our approach builds on the recent progress in applying Deep Learning to molecule property predictions \cite{Wu2019,Zhang2022,Gilmer2017,molecules25010044}. But, in contrast to these works, our data set is rather small with only a few thousand compounds. To compensate for this, we combine Deep Learning for property prediction with physiological based pharmacokinetic (PBPK) models. PBPK models are well established mechanistic models, describing the kinetics of compounds in physiological environments \cite{Rodgers2005,Rodgers2006}. Doing so we benefit from our knowledge about rat physiology and the interplay of different processes, and make much more efficient use of the available data to learn relevant molecule characteristics.

\section{Methods and materials}\label{sec:methods}
In this section we first describe our hybrid model and give a brief overview on PBPK models. Then we describe the training procedure for our model. Finally, we give an overview over the data used to train our hybrid model

\subsection{Hybrid modelling} \label{sec:hybrid}
\begin{figure}[t]
\includegraphics[width=\textwidth,trim={0 8cm 0 2cm},clip=true]{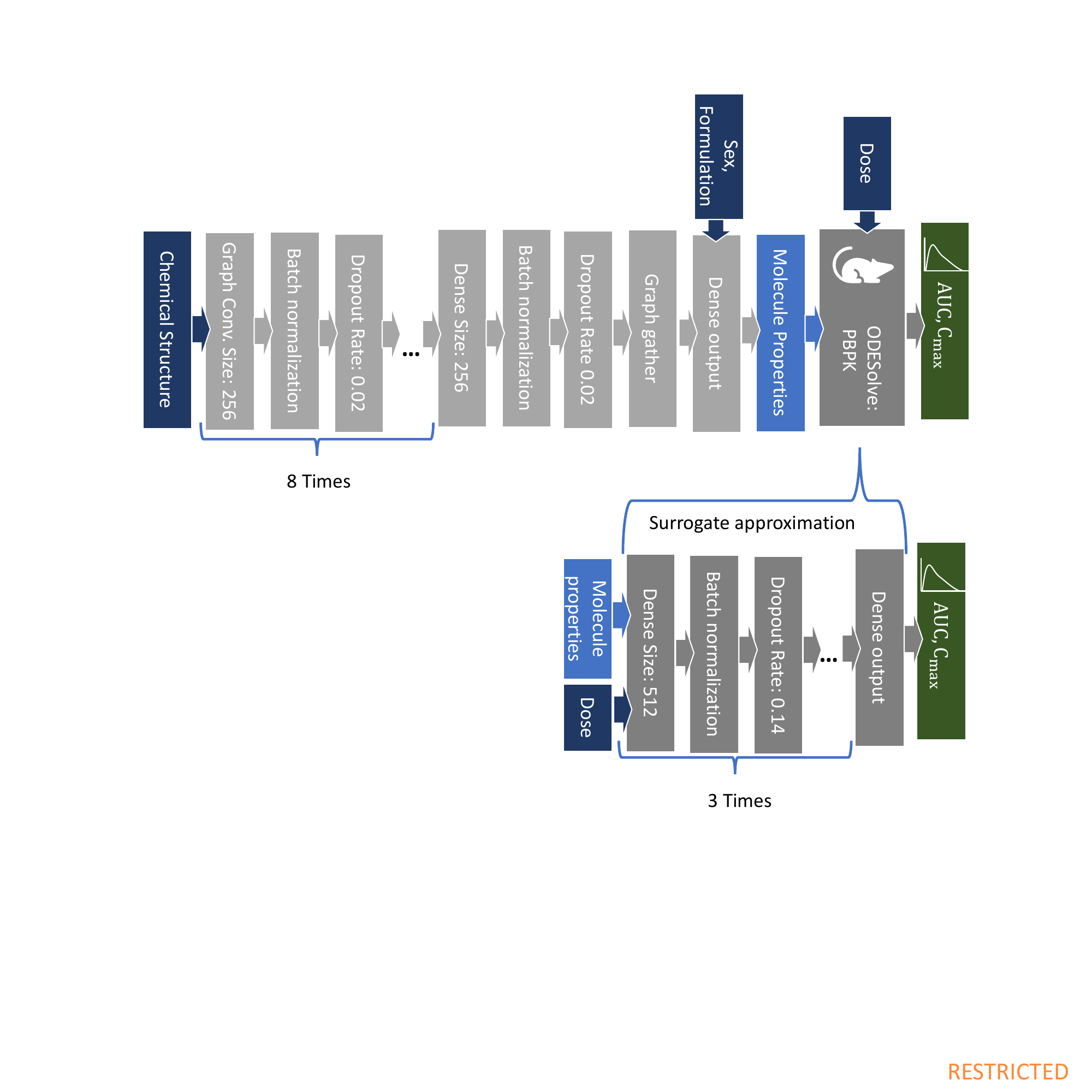}

\caption{Overview over our hybrid model structure consisting of a graph convolutional neural network for predicting a set of molecule properties. These molecule properties are the free parameters of a physiological model of rats predicting the pharmacokinetics. In practice, we approximate the PBPK model by a surrogate neural network.}
\label{fig:hybridmodel}
\end{figure}

To predict pharmacokinetics in rats we combine Deep Learning for molecular property prediction with PBPK models. A PBPK model is a system of ordinary differential equations (ODE) describing the PK processes a compound is undergoing within an organism. The processes are usually referred to as ADME processes, which stands for administration, distribution, metabolism and excretion. PBPK models are compartment models in which organs are represented by the compartments and the processes are parametrized by physicochemical and other molecule properties. For readers not familiar with PBPK models we provide a brief overview in \ref{ap:pbpk}.

In our approach, depicted in \figref{fig:hybridmodel}, the molecular properties predicted by a neural network, here called \blackboxname, correspond to parameters of the PBPK model, e.g. the solubility and the amount of substance cleared in the liver (hepatic clearance). The other parameters of the PBPK models are compound independent and describe the physiology of the organism, e.g., the organ volumes, blood flows, or specify the drug administration, e.g., administration route (IV/PO), dose and formulation.

The clear split between parameters describing the physiology and the molecule in our model highlights the expected advantages of our hybrid model in terms of required data and generalization. In our model the physiological parameters are fixed by the choice of target organism, so the neural network bridges between the molecular structure and the physiology. Furthermore, as certain aspects of the problem, e.g. the dose dependency, are mechanistically modelled, our model is able to exploit the extrapolation capabilities of the PBPK model, for example by generalizing to dosages outside the training range or predicting properties it has not been trained on, e.g. concentrations in different tissues.

Furthermore, we can exploit the flexibility of the \blackboxname{ }to compensate for a misspecified or inaccurate mechanistic model. Two examples we consider in this publication, are the differences between male and female rats and different formulations used for the oral administration. We model those cases by a mechanistic model for male rats and solution as formulation, but let the molecule properties depend on these covariates by passing them to the (last layers) of the \blackboxname. By doing so, the \blackboxname{ }learns to adapt its outputs such that even though we are using a wrong mechanistic model, we are still able to obtain an accurate model for female rats and suspensions.
\begin{table}[t]
 \begin{tabularx}{1.0\textwidth}{ 
    | >{\raggedright\arraybackslash}X 
    | >{\raggedright\arraybackslash}X 
    | >{\raggedright\arraybackslash}X 
    | >{\raggedright\arraybackslash}X 
    | >{\raggedright\arraybackslash}X |}
    
    \hline
    \textbf{Parameter} & \textbf{Short description} & \textbf{Pretraining data source}& \textbf{Distribution for surrogate model training}\\
    \hline
    Hepatic clearance& Clearance rate in liver& In-vitro data (from Hepatocyte stability assay) & Log-normal\\
    \hline
    $\rm{V_{max}}$& P-gp like transporter& In-vitro data (Caco-2 assay) & Mixtrure of point mass at 0 and a half-normal\\
    \hline
    GFR fraction& Fraction filtered in kidney & No pretraining & Uniform\\
    \hline
    Fraction unbound& Fraction unbound in plasma & Predicted by independent DNN &Truncated normal\\
    \hline
    Lipophilicity& Membrane affinity (log(MA))& Predicted by independent DNN & Normal\\
    \hline
    Effective molecular weight& Surrogate for molecule size & Calculated molecular weight excluding halogens& Half-normal\\
    \hline
    Stomach solubility& Solubility in stomach& Predicted using Henderson-Hasselbach equation, DNN for solubility and pKa&Log-normal\\
    \hline
    Small intestine solubility& Solubility in small intestine& Predicted using Henderson-Hasselbach equation, DNN  for solubility and pKa&Log-normal\\
    \hline
    Large intestine solubility& Solubility in large intestine& Predicted using Henderson-Hasselbach equation, DNN for solubility and pKa&Log-normal\\
    \hline
    Small intestine permeation& Absorption rate in small intestine&  Predicted from predicted log(MA) and molecular weight&Log-normal\\
    \hline
    Large intestine permeation&  Absorption rate in large intestine& Predicted from predicted log(MA) and molecular weight&Log-normal\\
    \hline
\end{tabularx}
\caption{Overview over the compound properties used as input of the PBPK model and whether predicted or in-vitro observed values are used for pertaining. Details on the used prediction models can be found in \cite{Goeller2020}.}
\label{tab:physchem}
\end{table}

\subsection{Physiologically based pharmacokinetic models}\label{sec:pbpk}

For the mechanistic model, we use the generic rat PBPK model available in the Open Systems Pharmacology (OSP) Suite \cite{OSPS2019} and add a generic hepatic (metabolism) and renal clearance (glomerular filtration in kidney) as well as a generic global P-glycoprotein (P-gp)-like active transport, which causes a flow from the inside to the outside of cells. For our purpose of predicting bioavailability in the early phases of drug and agrochemical development, it is sufficient to fix the physiological parameters to those of a typical rat, by using the OSP default values. The compound properties used as input for the PBPK model are listed in \tabref{tab:physchem}. For oral administrations we assume a solution as formulation and account for differences between solution and suspension as well as differences between male and female rats by passing formulation and sex to the \blackboxname{ }as described in \secref{sec:hybrid}.

\subsection{Neural network architecture}\label{sec:NNarchitecture}

Compared to the model developed in \cite{Schneckener2019} we here replace the SMILES string representation of molecules and the corresponding 1D convolutional architecture with a graph convolutional network (GCN) architecture directly acting on the graph representation of molecules. 
As SMILES representations are generated by a depth-first traversal through the molecular graph with an arbitrary starting node, they are not unique and connected sub-graphs are neither represented as contiguous sub-strings nor are they represented in same way when occurring in different molecules. 
In contrast,  graph convolutional layers explicitly respect permutation invariance of the graph nodes and the connectivity of the graph. Hence, they do not suffer from the non-uniqueness and connectivity issues of SMILES based architectures.
 We therefore expect a GCN architecture to be superior to a SMILES based architecture. Indeed, we even have difficulties finding a set of hyperparameters for the SMILES based architecture, which result in stable pre-training and acceptable accuracy, while we easily found those hyperparameters for the GCN.

We use the GCN-architecture proposed in \cite{Duvenaud2015} implemented in deepchem \cite{Ramsundar2019}. This GCN-architecture uses differentiable operations inspired by those used to calculate circular fingerprints and equips them with learnable weights.

\subsection{Model training}\label{sec:training}
\subsubsection{Surrogate}\label{sec:surrogate}
\begin{figure}[t]
    \includegraphics[width=0.32\textwidth]{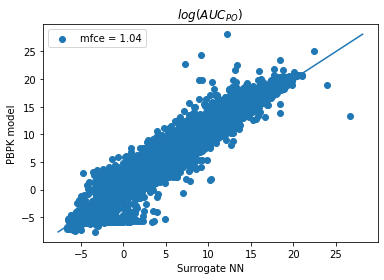}
    \includegraphics[width=0.32\textwidth]{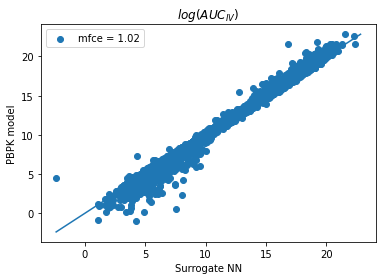}
    \includegraphics[width=0.32\textwidth]{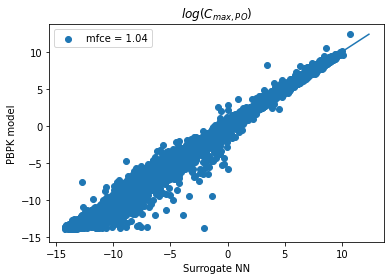}
    \caption{Overall accuracy of surrogate neural evaluated on a hold out test set of simulations. A median fold change error of 2-4\% is small to the expected biological variability of the data of ~50\%.}
    \label{fig:surrogateValidation}
\end{figure}
\begin{figure}%[t]
    \centering
    \includegraphics[width=0.49\textwidth]{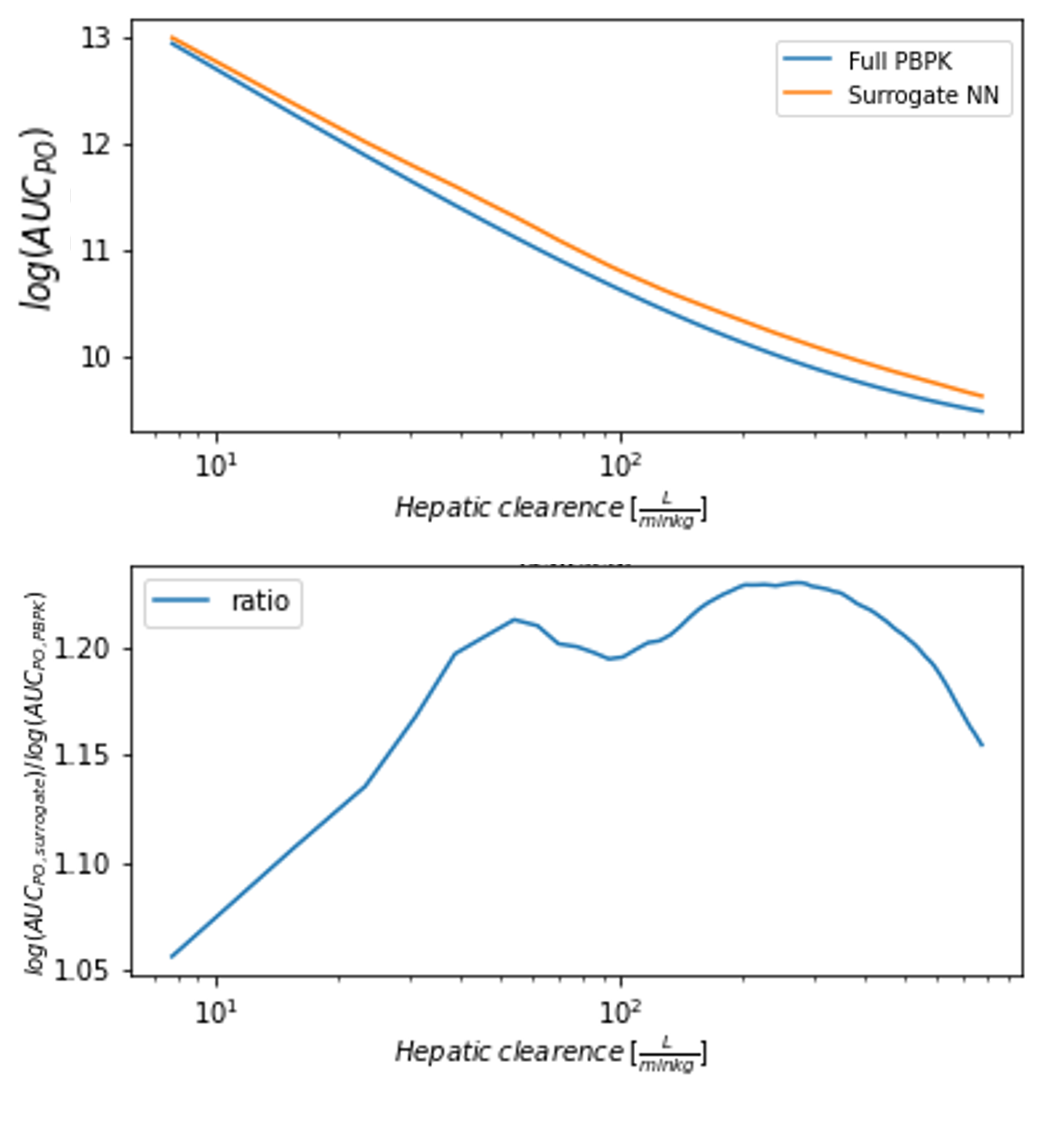}
    \includegraphics[width=0.49\textwidth]{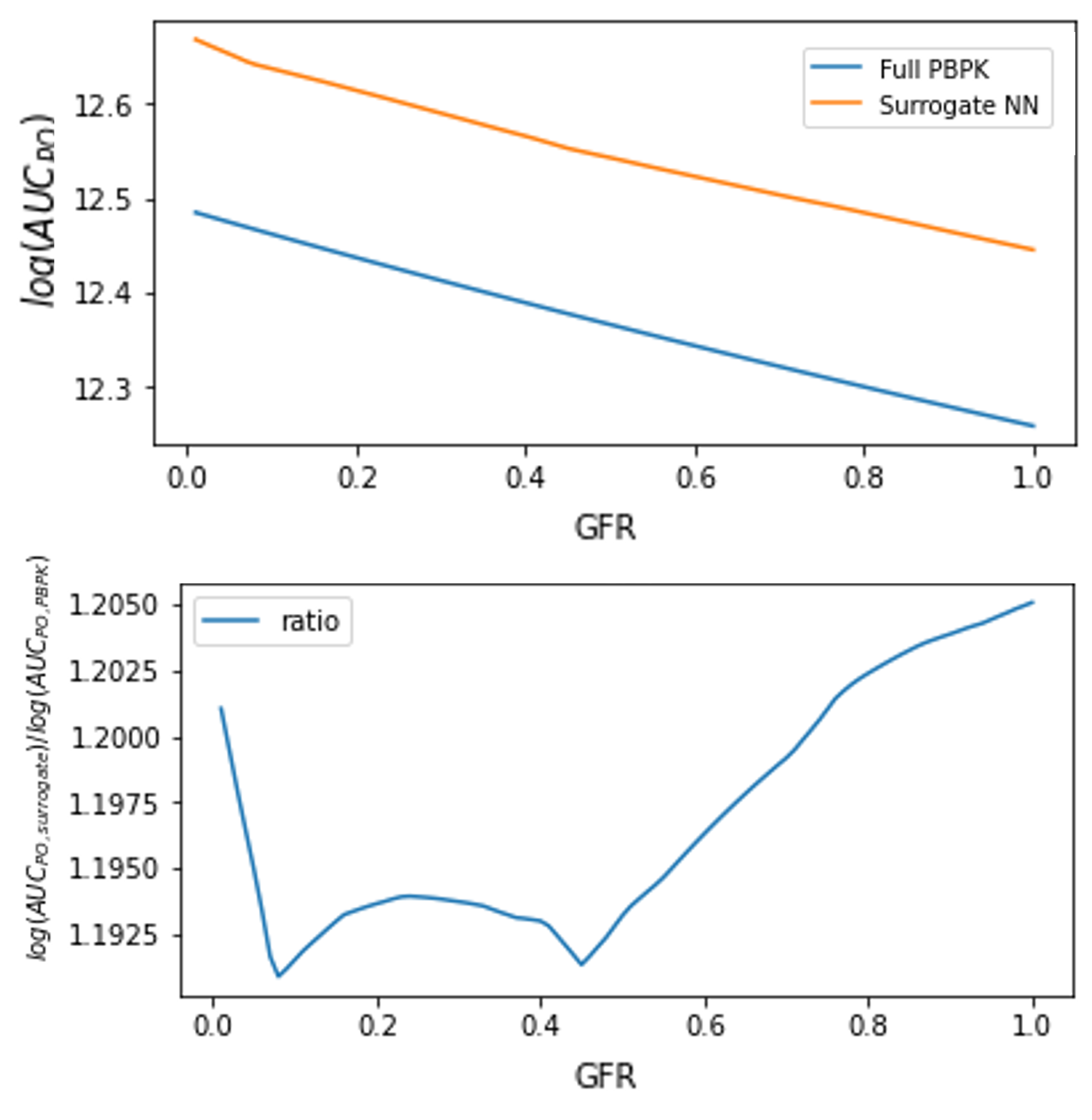}\\
    \includegraphics[width=0.49\textwidth]{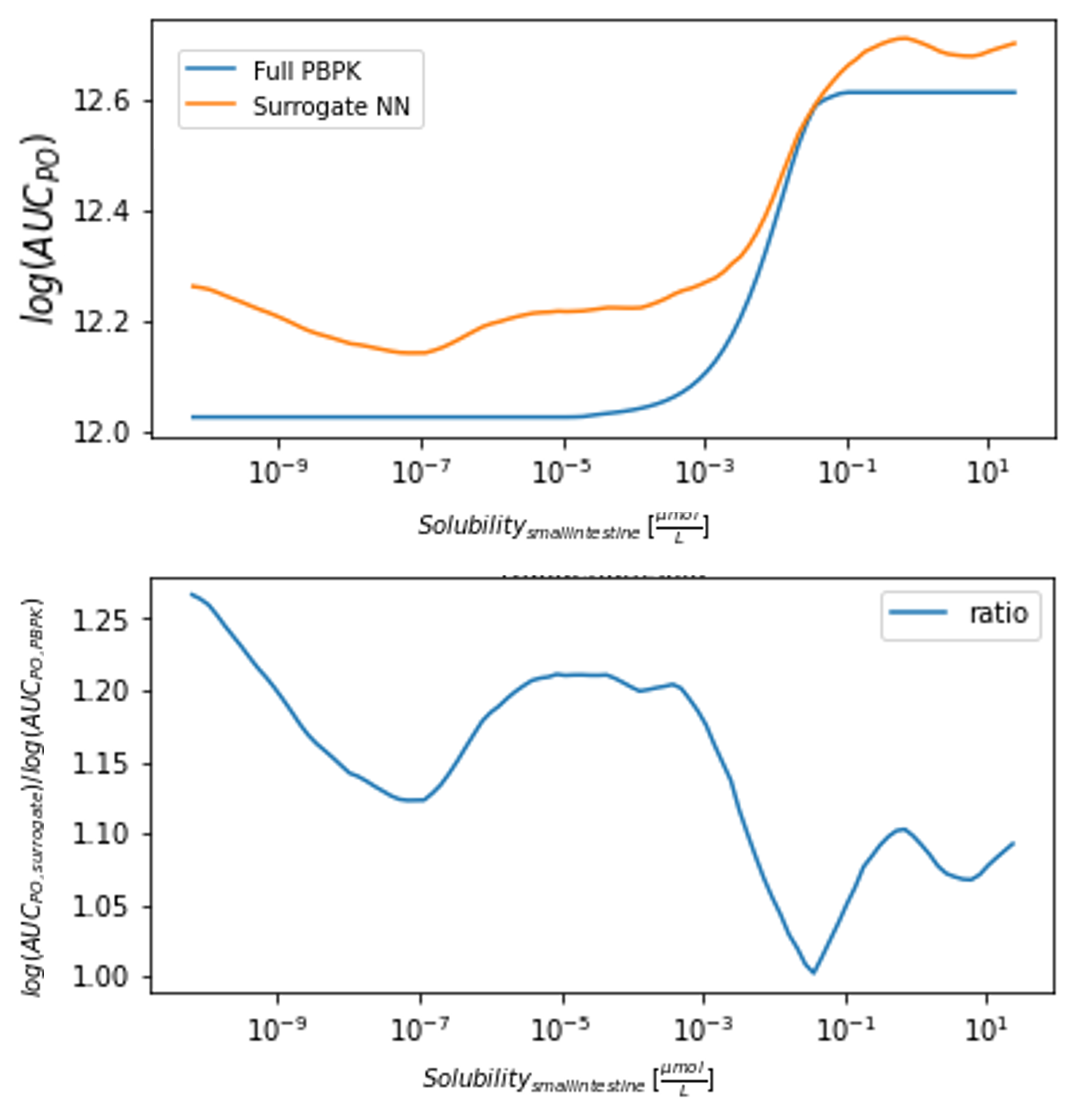}
    \includegraphics[width=0.49\textwidth]{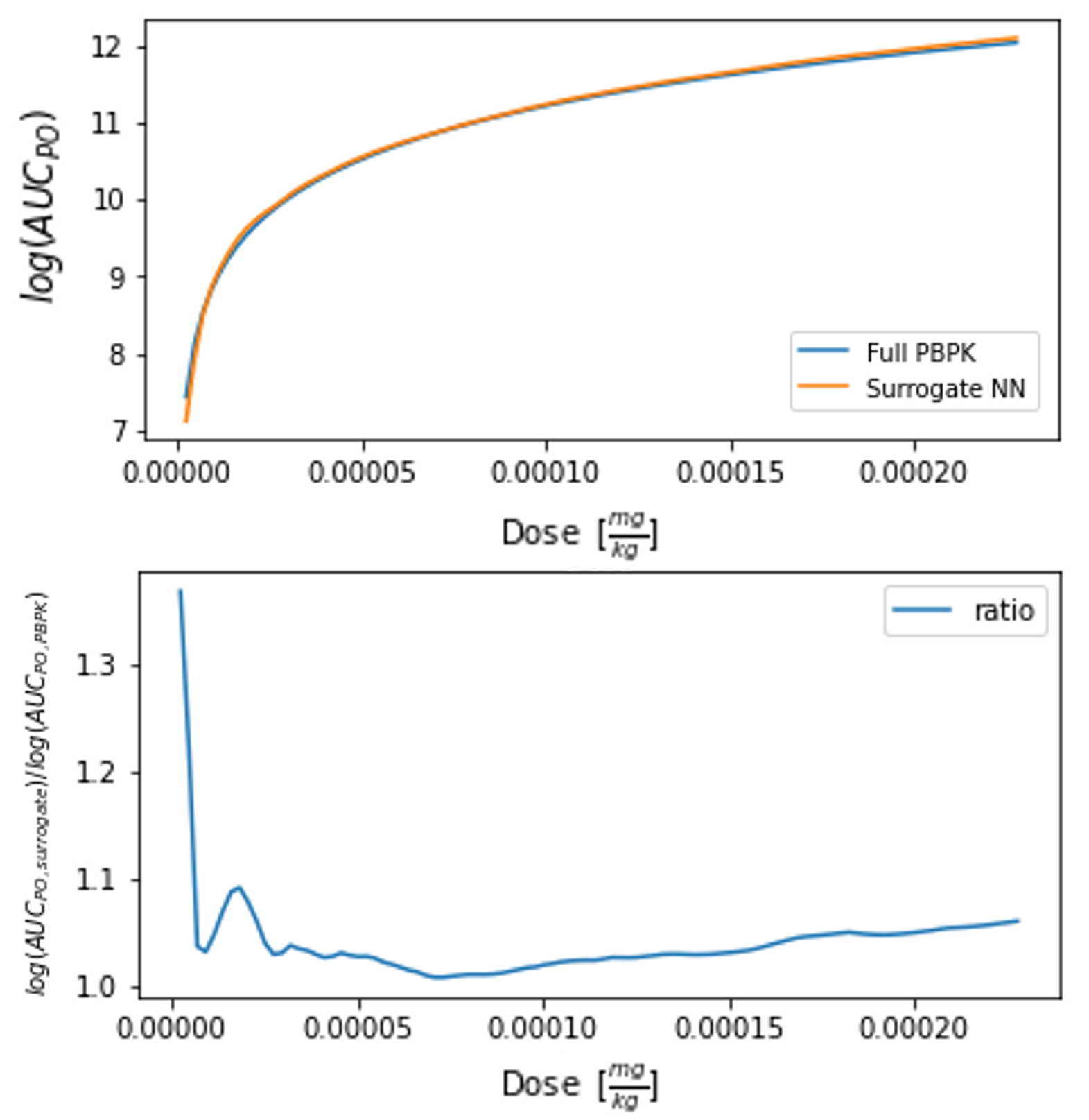}
    \caption{Some examples showing the full PBPK model and the surrogate model as a function of a single model parameter, while keeping the others fixed. In these examples the dependence on the hepatic clearance (top left) and dose (bottom right) is very accurately described by the surrogate. In the GFR example (top right) the surrogate is able to reproduce the shape of the PBPK, but shows a constant offset of about 20\%, which is acceptable given the variability of the PK-data. The solubility example (bottom left) shows an offset of similar size, but is able to qualitatively reproduce the step-like behavior seen in the PBPK model. The small oscillations of about few \% do not introduce major problems during training of the hybrid model.}
    \label{fig:surrogateValidationDependence}
\end{figure}\
For end-to-end training of the hybrid model, we need to back-propagate through the PBPK model. Even though this is possible for small ODE systems, it is computationally prohibitive for our model with about 300 stiff ODEs. Therefore, we replace the PBPK model by a surrogate neural network approximating the PBPK model. Here, we use a fully connected neural network. We train the surrogate model on 2.4M simulations with random model parameters, sampled using latin hyper-cube sampling, and test it on additional 0.72M randomly sampled simulations.  
Each parameter is distributed according to a simple parametric distribution, e.g. normal or log-normal, roughly matching the distribution of values in our training data. The functional form of the distributions are summarized in \tabref{tab:physchem}. We increase the variances of the distributions by 50\,\% to avoid the predictions of the \blackboxname{ }leaving the training range of the surrogate. As can be seen from \figref{fig:surrogateValidation}, the surrogate is able to reproduce the PBPK model accurately, with a fold change error of about 1.04 for the two PO endpoints and 1.02 for the IV endpoint. The error of the surrogate is negligible compared to the expected error caused by the high biological variability of about 50\% of the in-vivo data.

To be able to reliably back-propagate through the surrogate, good point wise approximations of the PBPK model are not sufficient. Also, the surrogate gradients and ideally higher order derivatives need to be good approximations of the PBPK models derivatives, i.e. the surrogate needs to reproduce the response of the PBPK model to changes in the molecule properties. We confirm this qualitatively by randomly generating a set of points in the PBPK input space and then vary each molecule property individually, while holding the others fixed, some examples are shown in \figref{fig:surrogateValidationDependence}. Overall, we find good agreement between the curves predicted by the surrogate and the PBPK model. 

\subsubsection{Training strategy}\label{sec:trainingstrategy}
To overcome the small data set of about 7000 compounds, we pre-train the \blackboxname{ }on molecule properties of about 140k compounds. sing seemingly unrelated data and targets is often sufficient for pre-training \cite{Ackermann2018,Kim2022,Farrens2022}, we therefore use predicted properties where available and measured values otherwise. Table \ref{tab:physchem} indicates for which properties predicted and for which measured values are used. Details on the pretraining data set are given in \secref{sec:data}.
The pretrained \blackboxname{ }is then trained end-to-end, as part of the hybrid model, on the in-vivo data to predict the target PK-parameters.

To constrain the model parameters predicted by the \blackboxname{ }to physiological values, e.g. a non-negative clearance, and to the range of the surrogate training data, we add a penalty term to the loss function $L_{total}$ 
\begin{equation}
L_{total}=\sum_{i}{L\left(y_i,\widehat{y_i}\right)}+\lambda\sum_{i,j}{max\left(p_{j,i}-p_{max,j},0\right)^2+max\left(p_{min,j}-p_{j,i},0\right)^2}.
\label{eq:loss}
\end{equation}
The first sum is over all data points $i$, the second over all data points $i$ and all molecule properties indexed by $j$. $p_{i,j}$ denotes all predicted molecule properties for all data points. The penalty is zero as long as $p_{min,j}\leq p_{i,j} \leq p_{max,j}$ and positive otherwise.
When during training the penalty term is not decreasing for several epochs, we increase the weight $\lambda$ until a pre-defined tolerance, here $10^{-8}$, is reached. Empirically, we find that this is sufficient to constrain the PBPK model parameters to the viable range, see \secref{ap:pcrange}. 

\subsection{Data}\label{sec:data}

\begin{figure}[t]
    \centering
    \includegraphics[width=0.59\textwidth]{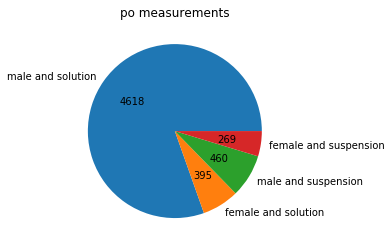}
    \includegraphics[width=0.39\textwidth]{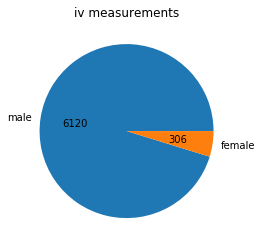}
    \caption{Number of data points for different sub-sets of the data set. Since the standard test for pharmaceuticals is on male rats using, for PO, a solution, most of our compounds are tested on male rats.}
    \label{fig:dataoverview}
\end{figure}
\begin{figure}[t]
    \centering
    \includegraphics[width=0.49\textwidth]{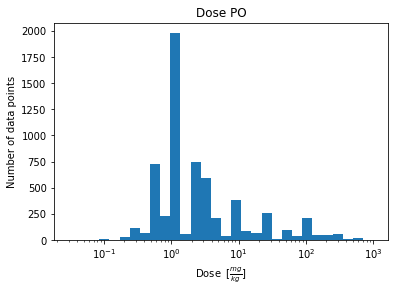}
    \includegraphics[width=0.49\textwidth]{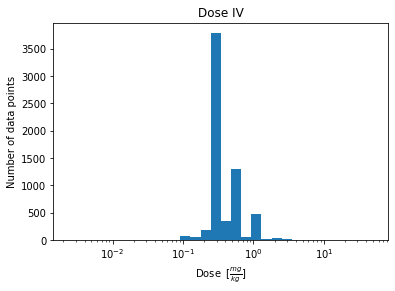}
    \caption{Distribution of used dose in PO (left) and IV (right) measurements. High doses are typically tested only in PO experiments, hence they span much large dose range then the iv experiments.}
    \label{fig:dosedistribution}
\end{figure}

We retrieved all in-vivo data taken after PO or IV administration in Wistar rats from the Bayer data warehouse. After filtering out pro-drugs, salts, molecules heavier than $1500 \, g / mol$ and non-standard formulations we are left with 7192 compounds, with in total 5731 $AUC_{PO}$, 6183 $C_{max,PO}$ and 6408 $AUC_{IV}$ measurements. In contrast to \cite{Schneckener2019}, in
 addition to male rats and PO administrations as solution we consider also female rats and suspensions. Furthermore, in contrast to our previous work \cite{Schneckener2019}, we do not restrict the dose, therefore, our data covers a dose range from $0.0024 \,\, mg / kg$ to $1000 \, mg / kg$.  Overall, there is more data available for low dosages than for high dosages, see \figref{fig:dosedistribution}. The large dose range reflects the fact that our data set includes relative low dose ($\sim 1 \, mg / kg$) data mainly taken from male rats at Bayer Pharmaceuticals, as well as high dose data  ($\gtrsim 10 \, mg / kg$)  mainly taken from female rats at Bayer CropScience. Furthermore, the compounds from both divisions are expected to have different properties. This increases the diversity in our data set. We expect that this results in a better generalization of the model. To challenge the capabilities of our hybrid model to generalize to new observables we also collect available \AUCday data. We select 20\% of the compounds randomly for testing the models performance. This test set is not used for training the model.

 For pretraining, we use about 100k compounds from the Bayer data warehouse. We use our internally available models to predict solubility, pKa values, lipophilicity and plasma protein binding in human for all compounds. For the hepatic clearance and membrane permeation no model is available, so we use all available in-vitro measurements resulting in an additional 40k compounds for pretraining. Note that usually no urine data is collected, so neither data nor a model is available for the GFR fraction, hence the GFR fraction is only trained end-to-end. In total, we use about 140k compounds for pre-training. We ensured that none of the compounds used for pretraining the model is part of our in-vivo data set.

\section{Results}\label{sec:results}
In this section we validate the predictive performance of our model and compare it to a standard GCN. We further challenge the generalization capabilities of our hybrid model by using it to predict the $\rm{AUC_{24h}}$, a quantity the model has not been trained on.

\subsection{Model performance}\label{sec:performance}
\begin{figure}[t]
    \centering
    \includegraphics[width=0.49\textwidth]{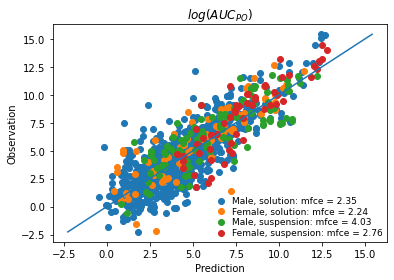}  
    \includegraphics[width=0.49\textwidth]{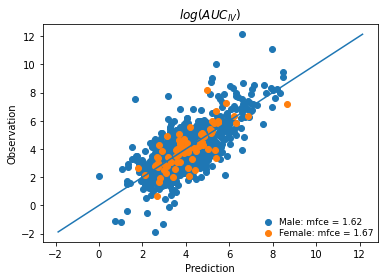}\\
    \includegraphics[width=0.49\textwidth]{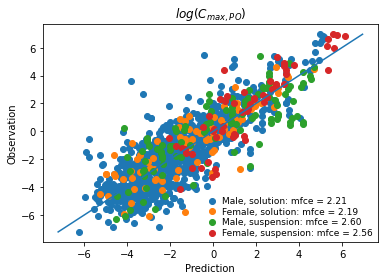}
    \includegraphics[width=0.49\textwidth]{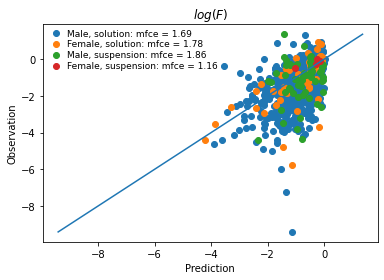}
    \caption{Model predictions vs observed values for our three training tasks \AUCPO (top left), \AUCIV (top right) and \Cmax (bottom left). For comparison with earlier work \cite{Schneckener2019,Daga2018} we also show the derived predictions for $F$ (bottom). The predictions for \AUCIV are more accurate than the PO predictions, this is expected since the processes involved in a PO administration are more complex than those involved in an IV administration. We observe that despite having less data for female rats than for male rats the predictions have a similar accuracy, while predictions for suspensions are a bit less accurate than those for solutions.}
    \label{fig:overallvalidation}
\end{figure}

We optimize the hyperparameters of our hybrid model using the HORD algorithm \cite{Ilievski2016}. The model architecture is optimized on the pretraining set while training hyperparameters are optimized on the training set. We validate the best model on our 20\% hold out test set in \figref{fig:overallvalidation}. For evaluation, we use the median fold change error defined as:
\begin{equation}
mfce=\ exp\left(median\left|log\left(y\right)-log\left(\hat{y}\right)\right|\right),
\label{eq:mfce}
\end{equation}
such that for a perfect model $mfce=1$. A fold change error of 2 to 3 is considered sufficient to inform compound selection \cite{Naga2022,Jones2015,Buck2007,Rowland2011a,Maharaj2014,Jones2011,Wang2019}. For all targets, except for the \AUCPO for male rats and suspension, we reach this goal. For \AUCIV and $F$ our model even achieves $mfce<2$. Compared to previous work \cite{Schneckener2019}, which uses a slightly different test set, the $mfce$ of AUC predictions for male rats for PO (solution) and IV has improved from $2.85$ to $2.35$ (PO) and from $1.95$ to $1.62$ (IV). Also the $mfce$ of $F$ predictions improved from $1.83$ to $1.62$. 
Additionally, we observe a more stable training and easier to tune hyperparameters. \Cmax after oral administration, which has not been considered in \cite{Schneckener2019}, can be predicted with a slightly higher accuracy than the AUC. 
Even though there is less data available for female rats than for male rats, predictions for female rats after an IV administration or a PO administration using solution can be made with similar accuracy as for male rats. We observe that PO predictions for suspensions are less accurate than predictions for solutions. This is expected, given that the dissolution of a suspension adds complexity to the dynamics in the GIT. Likewise, predictions for IV are more accurate than predictions for PO. As described in \secref{sec:hybrid} we predict suspension by using a mechanistic model for solutions with adapted molecule properties, we expect that using a mechanistic model for suspension, the prediction accuracy for suspension can be improved.

\subsection{Advantages of hybrid models}\label{sec:hybridvalidation}

\begin{figure}[t]
    \centering
    \includegraphics[width=0.32\textwidth]{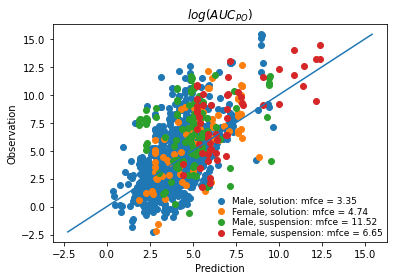}
    \includegraphics[width=0.32\textwidth]{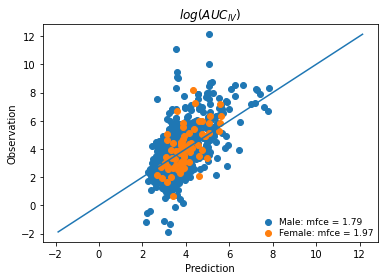}
    \includegraphics[width=0.32\textwidth]{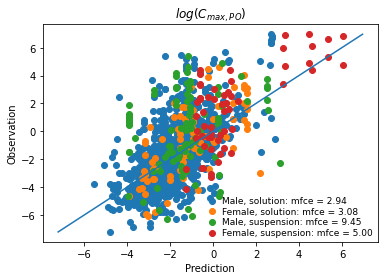}
    \caption{Model accuracy for a pure GCN model for the same three predictions tasks \AUCPO  (top), \AUCIV  (mid) and \Cmax  (bottom) as for the hybrid model. The accuracy of all 3 end-points is higher for the hybrid model than for the pure deep learning model, see \figref{fig:surrogateValidation}. }
    \label{fig:blackboxvalidation}
\end{figure}
\begin{figure}[t]
    \centering
    \includegraphics[width=0.49\textwidth]{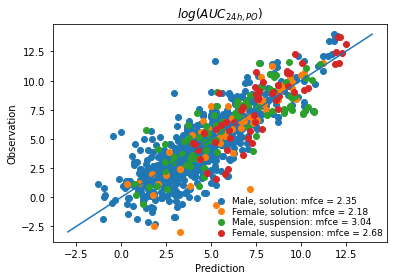}
    \includegraphics[width=0.49\textwidth]{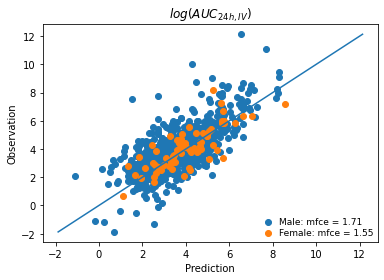}
    \caption{Predicted \AUCday compared to the observed values. The \AUCday are predicted from the molecule properties predicted by the \blackboxname, using the exact PBPK model. Despite not being trained on the \AUCday our hybrid model achieves an accuracy comparable to the total $AUC$.}
    \label{fig:24hvalidation}
\end{figure}

Figure \ref{fig:blackboxvalidation} shows for comparison the performance of a standard GCN, having the same architecture as the \blackboxname{ }except for the output layer's size. The predictions of the GCN are for all three endpoints worse than those of the hybrid model. While the performance drop of the \AUCIV predictions is moderate, the performance drop of \AUCPO  and \Cmax is of practical relevance as the standard GCN does not reach $mfce<3$.

In addition to the improved performance of the hybrid model, compared to a pure Deep Learning model, we can expect that the hybrid model is able to extrapolate and predict target parameters on which it has not been trained. For a first assessment of the extrapolation capability of our hybrid model we use the $\rm{AUC_{24h}}$. Figure \ref{fig:24hvalidation} shows the \AUCday predictions of our hybrid model compared to the observed values. The accuracy of the \AUCday predictions are comparable to the endpoints the model was trained on. Note that the \AUCday are predicted using the full PBPK model instead of the surrogate. The high predictive accuracy reconfirms that our surrogate is an accurate approximation of the PBPK model. This is further confirmed by the predictions of our training targets using the PBPK model instead of the surrogate model in \secref{ap:surrogate}.

\section{Summary and conclusion}\label{sec:conclusion}
In this work we present a hybrid model to predict the pharmacokinetics of pharmaceutical and agrochemical compounds bioavailability in rats directly from chemical structure. As predicting in-vivo targets is challenging due to the complex non-linear interplay of many processes as well as the low amount and high variability of data. We tackled these challenges by combining expert knowledge about rat physiology and processes affecting pharmacokinetics with Deep Learning for molecular property predictions.

The work in \cite{Schneckener2019} was extended by using a GCN for the \blackboxname{ }and changing the parametrization of the mechanistic model which improves the interface between the neural network and the mechanistic model. Additionally, we increased the available training data set by less restrictive filtering of the data and the inclusion of additional training endpoints, such as \Cmax or predictions for female rats and suspensions. Furthermore, we used a larger internal data set for pre-training, which is expected to be better correlated to bioavailability than the previously used public data from the TOX21 challenge \cite{Mayr2016}. That lead to an improved performance of the model compared to \cite{Schneckener2019}. An interesting subject for a future publications would be to investigate whether different architectures, e.g. Transformer, or pre-training strategies, e.g. self-supervised training, as in \cite{Fabian2020} can improve the model even further.

For all except one end-point our model has an $mfce<3$. Our $AUC_{IV}$ and $F$ predictions even have an mfce $<2$.
This is expected to be accurate enough to inform decisions during early phases of drug discovery \cite{Naga2022,Jones2015,Buck2007,Rowland2011a,Maharaj2014,Jones2011,Wang2019}. Furthermore, our model enables the selection and prioritization of compounds which are directly optimized with respect to their pharmacokinetic profile \cite{GomezBombarelli2018,Winter2019,Winter2019a}.

 Our prediction accuracy is competitive to the $F$ prediction accuracy in \cite{Daga2018}, when their model is only trained on in-vivo data from the chemical series it is applied to and superior otherwise. Our model shows a similar performance to the different models in \cite{Obrezanova2022}, which seem to have a slightly higher accuracy, but to achieve this, the chemical structure as well as in-vitro parameters are required, whereas our approach does not rely on in-vitro measurements. We like to stress that such a comparison should not be over interpreted, as different data-sets are used for training and validating the different models.
 
 Additionally, our approach is able to handle different covariates like sex and formulation by predicting effective molecule properties, which are sex and formulation dependent. We expect that prediction the accuracy can be improved by accounting for these covariates in the mechanistic model. The inclusion of further covariates like body weight is therefore likely to result in even better predictions.  However, the incorporation of more covariates is limited in practice as the typically available covariates do not fully specify the physiology of an individual. To account for the residual variability and to possibly improve predictions further probabilistic models which estimate population parameter distributions are required. To do so one could build on the recent progress in building deep generative models \cite{Kingma2019,Papamakarios2019,Kobyzev2021}. 
 
 Incorporating more covariates - either deterministic or probabilistic - requires more input parameters to the mechanistic model, which complicates the use of a surrogate neural network for the PBPK model. Training and using a surrogate worked well for the small number of inputs and outputs considered in this work, but becomes harder if their number increases. In such cases the use of a full PBPK model can become superior. However, this is currently computationally infeasible. Recently, there has been increasing interest in combining differential equations and Deep Learning \cite{Rackauckas2020,Raissi2017,Raissi2017a} and, consequently, in tools to train these models \cite{Rackauckas2019,Merkelbach2022,Kidger2021}, such that one can expect using the full PBPK model will become feasible in the near future. A complementary approach is to use simpler and hence computationally cheaper PK models, such as compartmental models or a reduced version of the full PBPK model.

Using PBPK models directly would also alleviate the need for constraining the molecule properties to the validity range of the surrogate. Constraining the molecule properties by introducing a penalty term in the loss worked in our case, but still complicated model training. Using a PBPK model directly would also enable to train the hybrid model on concentration-time profiles, which would be highly desirable, since more accurate predictions of concentration time profiles would allow a much more detailed description of the pharmacokinetics of a compound.

Successful application of PK models not only depends on the prediction accuracy, but also on the possibility to estimate uncertainty on the prediction. Such, that decisions based on predictions are only made for molecules for which the model is expected to be accurate. Ref. \cite{Obrezanova2022} assesses prediction uncertainty. For none of the tested approaches the uncertainty estimates are fully satisfactory. But, approaches with statistical correct Bayesian or Frequentist epistemic uncertainty provide better uncertainty estimates. We expect that in both cases well calibrated uncertainties can be provided by computationally expensive ensembles techniques \cite{Ashukha2020,ValentinJospin2020,Gawlikowski2021}.

Our model has the potential to reduce cost, development time and animal experiments in drug and agrochemical research by focusing the development on the most promising candidates and being able to directly optimize a compounds PK.  Furthermore, our approach can be used to predict human PK \cite{Gruber2024}, therefore directly optimizing for clinical use.

\section*{Declaration of Competing Interest} The authors declare no competing financial interest.
\section*{Acknowledgments}

\begin{appendices}
\section{Physiologically based pharmacokinetic models}\label{ap:pbpk}
Physiologically based pharmacokinetic (PBPK) models are ordinary differential equation models describing how a substance, e.g. a drug, is absorbed, distributed, metabolized, and excreted in an organism. For the reader not familiar with PBPK models we provide a brief overview over the basic concepts, building blocks and equations forming a PBPK model. For more details we refer to \cite{Peters2012}.

In PBPK models physiological organs and tissues are represented by compartments. The transport of substance via the blood is modeled by balance equations of the form
\begin{equation}
\frac{dC_i}{dt}=\frac{Q_i}{V_i}\left(C_{art}-\frac{C_i}{P_i}\right),\label{eq:transport}
\end{equation}
where $C_i$ denotes the compound concentration in the compartment $i$, $V_i$ its volume, $Q_i$ the blood flow, $P_i$ the partition coefficient between blood and tissue, and $C_{art}$ the compound concentration in arterial blood, which is governed by
\begin{equation}
\frac{dC_{art}}{dt}=-\sum_{i}{\frac{Q_i}{V_i}\left(C_{art}-\frac{C_i}{P_i}\right)},\label{eq:transportArt}
\end{equation}

To describe dissolution, absorption, metabolism and excretion, as well as additional distribution mechanism the equations \ref{eq:transport} and \ref{eq:transportArt} need to be extended. For example, dissolution and absorption in a single GIT compartment is described by the following equations:
\begin{align}
\frac{dC_g}{dt}&=\frac{Q_g}{V_g}\left(C_{art}-\frac{C_g}{P_g}\right)+K_a C_{lum},\label{eq:absorption}\\
\frac{dC_{lum}}{dt}&=-K_aC_{lum}+\frac{dC_{dis}}{dt},\label{eq:dissolved}\\
\frac{dC_{dis}}{dt}&=K\left(C_0-C_{dis}\right)^{2/3}\left(C_s-C_{lum}\right),\label{eq:noyseWhitney}
\end{align}

Equation \ref{eq:absorption} describes concentration in the GIT tissue $C_g$, which is sourced by a linear absorption process from the GIT lumen. Equation \ref{eq:dissolved} is describes the compound concentration in the GIT lumen $C_{lum}$, which is sourced by the dissolved compound $C_{dis}$. Equation \ref{eq:noyseWhitney} is the Noyse-Withney equation describing the dissolution of the compound in the GIT lumen, with $K$ being a compound dependent constant, $C_0$ is the total amount of compound administered divided by the administered volume and $C_s$ is the solubility, i.e. the compound concentration the GIT lumen at (thermal) equilibrium.
Metabolism is described by the Michaelis-Menten-Kinetics, which for $C\ll K_m$ can be linearized:
\begin{equation}
\frac{dC}{dt}=-V_{max}\frac{C}{K_M+C}= -\frac{V_{max}}{K_M} C+O\left(\left(\frac{C}{K_m}\right)^2\right),
\end{equation}
The constants $V_{max}$ and $K_M$ depend on the compound and the metabolizing enzyme and control the speed and saturation of metabolism. We assume a single generic metabolizing enzyme, hence in our hybrid model hepatic clearance is fully characterized by the rate $\frac{V_{max}}{K_M}$. 

An active P-gp like transport via membrane proteins, assuming a constant protein concentration, follows also a Michaelis-Menten-Kinetics
\begin{align}
\frac{dC_1}{dt}&=-V_{max}\frac{C_1}{K_M+C_1}\\
\frac{dC_2}{dt}&=V_{max}\frac{C_1}{K_M+C_1}.
\end{align}
As for the metabolism, the constants $V_{max}$ and $K_M$ control the speed and saturation of the transport are  compound and are transport protein dependent. For our purpose it is sufficient to set $K_M=1\: \rm{\frac{\mu mol}{L}}$, i.e. use the OSP default value, hence the transport is parametrized by its maximal velocity $V_{max}$.  

\section{Validation of property constraints}\label{ap:pcrange}

\begin{figure}[t]
    \centering
    \includegraphics[width=0.32\textwidth]{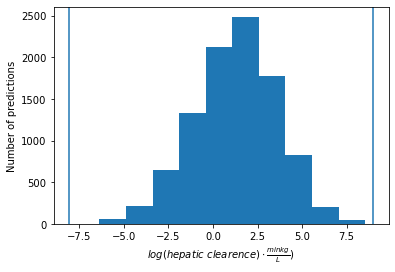}
    \includegraphics[width=0.32\textwidth]{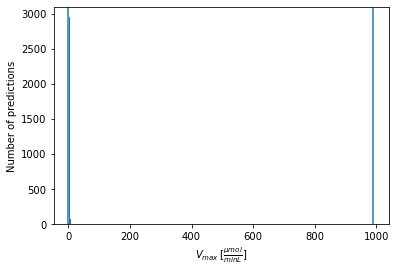}
    \includegraphics[width=0.32\textwidth]{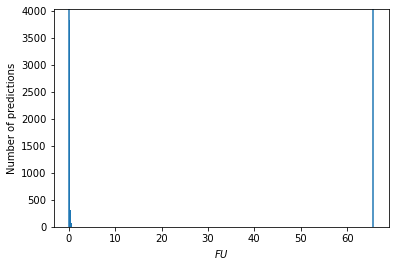}\\
    \includegraphics[width=0.32\textwidth]{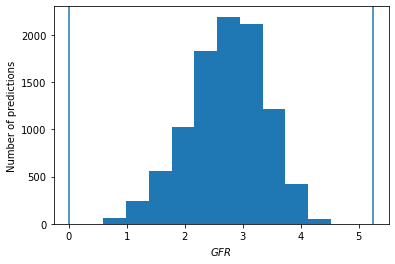}
    \includegraphics[width=0.32\textwidth]{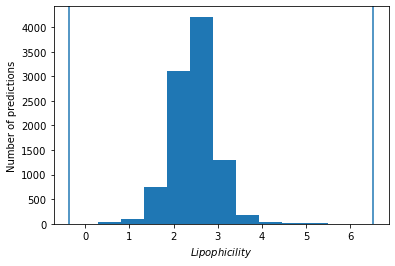}
    \includegraphics[width=0.32\textwidth]{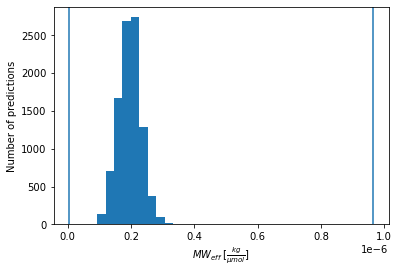}\\   
    \includegraphics[width=0.32\textwidth]{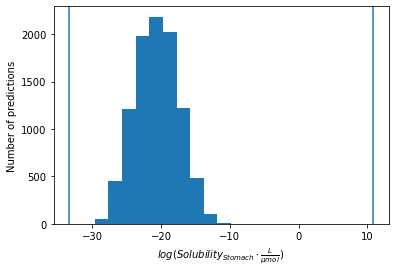}
    \includegraphics[width=0.32\textwidth]{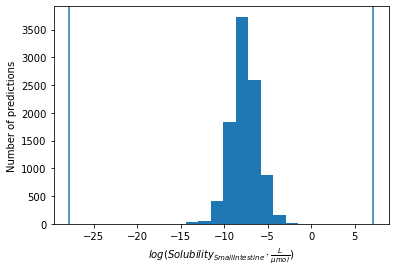}
    \includegraphics[width=0.32\textwidth]{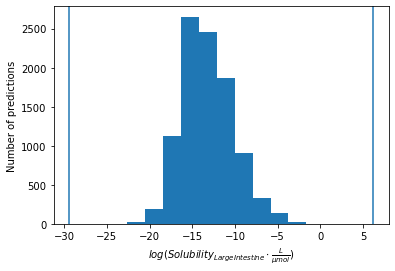}\\ 
    \includegraphics[width=0.32\textwidth]{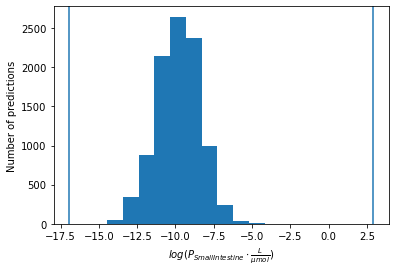}
    \includegraphics[width=0.32\textwidth]{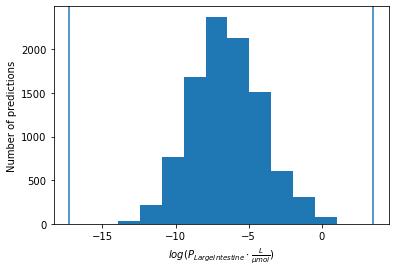}
    \caption{Distribution of the molecule properties in the test set. The vertical line show bounds for the properties to lie within in the validity range of the surrogate. All molecule properties  lie within in their bounds.}
    \label{fig:constraintvalidation}
\end{figure}
In \figref{fig:constraintvalidation} the distribution of predicted molecule properties of the test set are shown together with the maximal and minimal values in the surrogate training data set.  All predicted molecule properties lie within in the surrogates training range, confirming the effectiveness of the penalized loss described in \secref{sec:trainingstrategy}. Note that for $V_{max}$ and $FU$ we used heavy tailed distributions for generating the surrogate training data, resulting in the large range shown in \figref{fig:constraintvalidation}. For the $FU$ this results in unphysiological values $>1$, for which the equations of the PBPK model are still defined. But in practice the \blackboxname{ }does not predict a $FU >1$. Furthermore, to increase the flexibility of our clearance model we increased the maximal allowed value for the GFR fraction from 1 to 5.25.

\section{A posteriori surrogate validation}
\label{ap:surrogate}
\begin{figure}[t]
    \centering
    \includegraphics[width=0.32\textwidth]{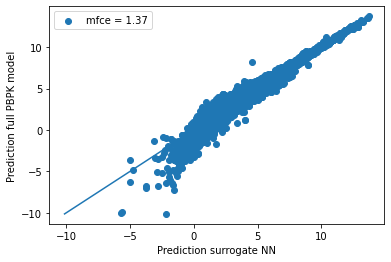}
    \includegraphics[width=0.32\textwidth]{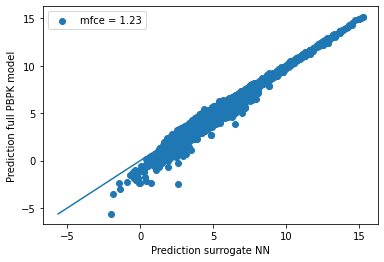}
    \includegraphics[width=0.32\textwidth]{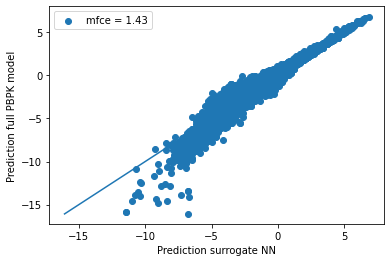}\\
    \caption{Simulation vs surrogate predictions for the predicted properties of the compounds in our test set for \AUCPO (left), \AUCIV (center) and \Cmax (right). The accuracy is a bit smaller compared to the estimate on the simulation test set, but still significantly better than the accuracy of the hybrid model, hence the accuracy of the surrogate is sufficient.}
    \label{fig:surrogateposthoh}
\end{figure}
\begin{figure}[t]
    \centering
    \includegraphics[width=0.32\textwidth]{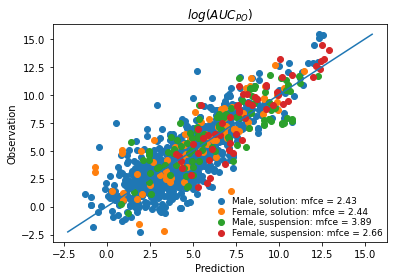}
    \includegraphics[width=0.32\textwidth]{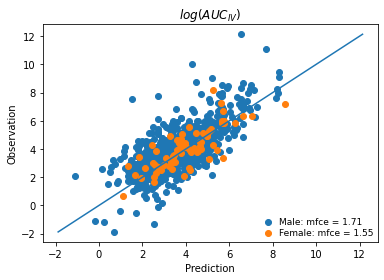}
    \includegraphics[width=0.32\textwidth]{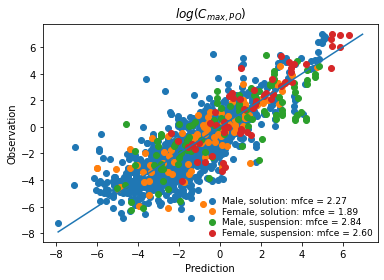}\\
    \caption{Hybrid model test set predictions using the full PBPK model instead of the surrogate predictions for the predicted properties of the compounds. The accuracy for the three end-points \AUCPO (left), \AUCIV (center) and \Cmax (right) is similar to the accuracy when using the surrogate. Demonstrating the accuracy of the surrogate model.}
    \label{fig:validationPBPK}
\end{figure}
We can validate the surrogate model a posteriori by predicting the training targets of our hybrid model using the PBPK model instead of the surrogate. Figure \ref{fig:surrogateposthoh} shows the predictions obtained using the PBPK model vs those obtained using the surrogate. The accuracy is not as good as expected from the analysis in \secref{sec:surrogate}, but still accurate enough to be used in the hybrid model, the $mfce$ of the surrogate ($1.2-1.4$) is clearly better than the $mfce$ of the hybrid model ($mfce\gtrsim1.6$). 
Additionally, \figref{fig:validationPBPK} shows the predictions using the full PBPK vs the observed values. These predictions are almost as accurate as those using the surrogate model. A maximal difference of 0.24 in the $mfce$ can be observed, and no additional features are visible. This highlights again the accuracy of the used surrogate model.

\section{Charge state dependence of model performance}
\label{ap:charge}
\begin{figure}[t]
    \centering
    \includegraphics[width=0.32\textwidth]{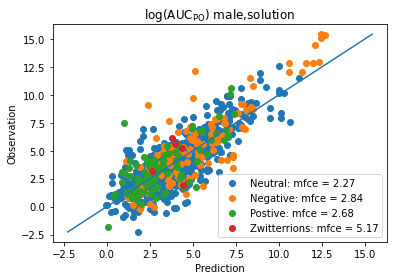}
    \includegraphics[width=0.32\textwidth]{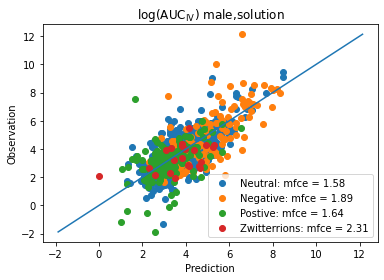}
    \includegraphics[width=0.32\textwidth]{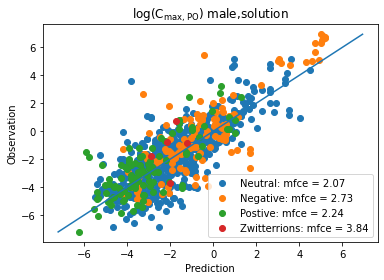}
    \caption{
    Dependence of the hybrid models accuracy on the compounds charge state at $pH=7.4$, i.e the pH value of blood. Shown are the three endpoints \AUCPO (left), \AUCIV (center) and \Cmax (right) for the case male rat and solution. For neutral compounds the predictions are most accurate, followed by positively and negatively charged compounds. For zwitterions the accuracy is significantly worse, but here the number of compounds is too low for a reliable estimate of the accuracy.}
    \label{fig:chargestate}
\end{figure}
We check for a potential dependence of the model accuracy on the charge state in \figref{fig:chargestate}. We evaluate the performance for male rats when a solution is used. As charge states can reliably be predicted, we use predicted charge states at the pH of blood ($pH=7.4$). We observe the best performance neutral compounds, and a worse performance for positively and negatively charged compounds. But, in all three cases we achieve $mfce<3$, so predictions are accurate enough to guide decisions. For zwitterions the mfce for \AUCPO and \Cmax is larger than $3$, but here only very few compounds are in our test set.
\end{appendices}
\bibliography{BA_ref}

\begin{thebibliography}{10}
\expandafter\ifx\csname url\endcsname\relax
  \def\url#1{\burl{#1}}\fi
\expandafter\ifx\csname urlprefix\endcsname\relax\def\urlprefix{URL }\fi
\providecommand{\bibinfo}[2]{#2}
\providecommand{\eprint}[2][]{\url{#2}}
\providecommand{\doi}[1]{\url{https://doi.org/#1}}
\bibcommenthead

\bibitem{Schneckener2019}
\bibinfo{author}{Schneckener, S.} \emph{et~al.}
\newblock \bibinfo{title}{Prediction of oral bioavailability in rats:
  Transferring insights from in vitro correlations to (deep) machine learning
  models using in silico model outputs and chemical structure parameters}.
\newblock \emph{\bibinfo{journal}{J. Chem. Inf. Model.}}
  \textbf{\bibinfo{volume}{59}}, \bibinfo{pages}{4893–4905}
  (\bibinfo{year}{2019}).

\bibitem{Tian2011}
\bibinfo{author}{Tian, S.}, \bibinfo{author}{Li, Y.}, \bibinfo{author}{Wang,
  J.}, \bibinfo{author}{Zhang, J.} \& \bibinfo{author}{Hou, T.}
\newblock \bibinfo{title}{Adme evaluation in drug discovery. 9. prediction of
  oral bioavailability in humans based on molecular properties and structural
  fingerprints}.
\newblock \emph{\bibinfo{journal}{Molecular pharmaceutics}}
  \textbf{\bibinfo{volume}{8}}, \bibinfo{pages}{841—851}
  (\bibinfo{year}{2011}).
\newblock \urlprefix\url{https://doi.org/10.1021/mp100444g}.

\bibitem{FalconCano2020}
\bibinfo{author}{Falc{\'o}n-Cano, G.}, \bibinfo{author}{Molina, C.} \&
  \bibinfo{author}{Cabrera-P{\'e}rez, M.~{\'A}.}
\newblock \bibinfo{title}{Adme prediction with knime: Development and
  validation of a publicly available workflow for the prediction of human oral
  bioavailability}.
\newblock \emph{\bibinfo{journal}{Journal of chemical information and
  modeling}}  (\bibinfo{year}{2020}).

\bibitem{Wu2019}
\bibinfo{author}{Wu, Z.} \emph{et~al.}
\newblock \bibinfo{title}{A comprehensive survey on graph neural networks}
  (\bibinfo{year}{2019}).

\bibitem{Zhang2022}
\bibinfo{author}{Zhang, Z.}, \bibinfo{author}{Cui, P.} \& \bibinfo{author}{Zhu,
  W.}
\newblock \bibinfo{title}{Deep learning on graphs: {A} survey}.
\newblock \emph{\bibinfo{journal}{{IEEE} Trans. Knowl. Data Eng.}}
  \textbf{\bibinfo{volume}{34}}, \bibinfo{pages}{249--270}
  (\bibinfo{year}{2022}).

\bibitem{Gilmer2017}
\bibinfo{author}{Gilmer, J.}, \bibinfo{author}{Schoenholz, S.~S.},
  \bibinfo{author}{Riley, P.~F.}, \bibinfo{author}{Vinyals, O.} \&
  \bibinfo{author}{Dahl, G.~E.}
\newblock \bibinfo{title}{Neural message passing for quantum chemistry}
  (\bibinfo{year}{2017}).

\bibitem{molecules25010044}
\bibinfo{author}{Montanari, F.}, \bibinfo{author}{Kuhnke, L.},
  \bibinfo{author}{Ter~Laak, A.} \& \bibinfo{author}{Clevert, D.-A.}
\newblock \bibinfo{title}{Modeling physico-chemical admet endpoints with
  multitask graph convolutional networks}.
\newblock \emph{\bibinfo{journal}{Molecules}} \textbf{\bibinfo{volume}{25}}
  (\bibinfo{year}{2020}).
\newblock \urlprefix\url{https://www.mdpi.com/1420-3049/25/1/44}.

\bibitem{Rodgers2005}
\bibinfo{author}{Rodgers, T.}, \bibinfo{author}{Leahy, D.} \&
  \bibinfo{author}{Rowland, M.}
\newblock \bibinfo{title}{Physiologically based pharmacokinetic modeling 1:
  Predicting the tissue distribution of moderate-to-strong bases}
  \textbf{\bibinfo{volume}{94}}, \bibinfo{pages}{1259--1276}
  (\bibinfo{year}{2005}).

\bibitem{Rodgers2006}
\bibinfo{author}{Rodgers, T.} \& \bibinfo{author}{Rowland, M.}
\newblock \bibinfo{title}{Physiologically based pharmacokinetic modelling 2:
  predicting the tissue distribution of acids, very weak bases, neutrals and
  zwitterions.}
\newblock \emph{\bibinfo{journal}{Journal of pharmaceutical sciences}}
  \textbf{\bibinfo{volume}{95}}, \bibinfo{pages}{1238--1257}
  (\bibinfo{year}{2006}).

\bibitem{Goeller2020}
\bibinfo{author}{Göller, A.~H.} \emph{et~al.}
\newblock \bibinfo{title}{Bayer’s in silico admet platform: a journey of
  machine learning over the past two decades} \textbf{\bibinfo{volume}{25}},
  \bibinfo{pages}{1702--1709} (\bibinfo{year}{2020}).

\bibitem{OSPS2019}
\bibinfo{title}{Open systems pharmacology suite}.
\newblock
  \urlprefix\url{https://github.com/Open-Systems-Pharmacology/Suite/releases/tag/v8.0}.

\bibitem{Duvenaud2015}
\bibinfo{author}{Duvenaud, D.} \emph{et~al.}
\newblock \bibinfo{title}{Convolutional networks on graphs for learning
  molecular fingerprints}  (\bibinfo{year}{2015}).
\newblock \urlprefix\url{http://arxiv.org/abs/1509.09292}.
\newblock \bibinfo{note}{Comment: 9 pages, 5 figures. To appear in Neural
  Information Processing Systems (NIPS)}.

\bibitem{Ramsundar2019}
\bibinfo{author}{Ramsundar, B.} \emph{et~al.}
\newblock \emph{\bibinfo{title}{Deep Learning for the Life Sciences}}
  (\bibinfo{publisher}{O'Reilly Media}, \bibinfo{year}{2019}).
\newblock
  \bibinfo{note}{\url{https://www.amazon.com/Deep-Learning-Life-Sciences-Microscopy/dp/1492039837}}.

\bibitem{Ackermann2018}
\bibinfo{author}{Ackermann, S.}, \bibinfo{author}{Schawinski, K.},
  \bibinfo{author}{Zhang, C.}, \bibinfo{author}{Weigel, A.~K.} \&
  \bibinfo{author}{Turp, M.~D.}
\newblock \bibinfo{title}{Using transfer learning to detect galaxy mergers}
  (\bibinfo{year}{2018}).

\bibitem{Kim2022}
\bibinfo{author}{Kim, H.~E.} \emph{et~al.}
\newblock \bibinfo{title}{Transfer learning for medical image classification: a
  literature review.}
\newblock \emph{\bibinfo{journal}{BMC medical imaging}}
  \textbf{\bibinfo{volume}{22}}, \bibinfo{pages}{69} (\bibinfo{year}{2022}).

\bibitem{Farrens2022}
\bibinfo{author}{Farrens, S.}, \bibinfo{author}{Lacan, A.},
  \bibinfo{author}{Guinot, A.} \& \bibinfo{author}{Vitorelli, A.~Z.}
\newblock \bibinfo{title}{Deep transfer learning for blended source
  identification in galaxy survey data} \textbf{\bibinfo{volume}{657}},
  \bibinfo{pages}{A98} (\bibinfo{year}{2022}).

\bibitem{Daga2018}
\bibinfo{author}{Daga, P.~R.}, \bibinfo{author}{Bolger, M.~B.},
  \bibinfo{author}{Haworth, I.~S.}, \bibinfo{author}{Clark, R.~D.} \&
  \bibinfo{author}{Martin, E.~J.}
\newblock \bibinfo{title}{Physiologically based pharmacokinetic modeling in
  lead optimization. 1. evaluation and adaptation of gastroplus to predict
  bioavailability of medchem series}.
\newblock \emph{\bibinfo{journal}{Molecular Pharmaceutics}}
  \textbf{\bibinfo{volume}{15}}, \bibinfo{pages}{821--830}
  (\bibinfo{year}{2018}).
\newblock \urlprefix\url{https://doi.org/10.1021/acs.molpharmaceut.7b00972}.
\newblock \bibinfo{note}{PMID: 29337578}.

\bibitem{Ilievski2016}
\bibinfo{author}{Ilievski, I.}, \bibinfo{author}{Akhtar, T.},
  \bibinfo{author}{Feng, J.} \& \bibinfo{author}{Shoemaker, C.~A.}
\newblock \bibinfo{title}{Efficient hyperparameter optimization of deep
  learning algorithms using deterministic rbf surrogates}
  (\bibinfo{year}{2016}).

\bibitem{Naga2022}
\bibinfo{author}{Naga, D.}, \bibinfo{author}{Parrott, N.},
  \bibinfo{author}{Ecker, G.~F.} \& \bibinfo{author}{Olivares-Morales, A.}
\newblock \bibinfo{title}{Evaluation of the success of high-throughput
  physiologically based pharmacokinetic ({HT-PBPK}) modeling predictions to
  inform early drug discovery}.
\newblock \emph{\bibinfo{journal}{Mol. Pharm.}} \textbf{\bibinfo{volume}{19}},
  \bibinfo{pages}{2203--2216} (\bibinfo{year}{2022}).

\bibitem{Jones2015}
\bibinfo{author}{Jones, H.~M.} \emph{et~al.}
\newblock \bibinfo{title}{Physiologically based pharmacokinetic modeling in
  drug discovery and development: a pharmaceutical industry perspective.}
\newblock \emph{\bibinfo{journal}{Clinical pharmacology and therapeutics}}
  \textbf{\bibinfo{volume}{97}}, \bibinfo{pages}{247--262}
  (\bibinfo{year}{2015}).

\bibitem{Buck2007}
\bibinfo{author}{Buck, S. S.~D.} \emph{et~al.}
\newblock \bibinfo{title}{Prediction of human pharmacokinetics using
  physiologically based modeling: A retrospective analysis of 26 clinically
  tested drugs} \textbf{\bibinfo{volume}{35}}, \bibinfo{pages}{1766--1780}
  (\bibinfo{year}{2007}).

\bibitem{Rowland2011a}
\bibinfo{author}{Rowland, M.}, \bibinfo{author}{Peck, C.} \&
  \bibinfo{author}{Tucker, G.}
\newblock \bibinfo{title}{{{P}hysiologically-based pharmacokinetics in drug
  development and regulatory science}}.
\newblock \emph{\bibinfo{journal}{Annu Rev Pharmacol Toxicol}}
  \textbf{\bibinfo{volume}{51}}, \bibinfo{pages}{45--73}
  (\bibinfo{year}{2011}).

\bibitem{Maharaj2014}
\bibinfo{author}{Maharaj, A.~R.} \& \bibinfo{author}{Edginton, A.~N.}
\newblock \bibinfo{title}{Physiologically based pharmacokinetic modeling and
  simulation in pediatric drug development}.
\newblock \emph{\bibinfo{journal}{CPT: Pharmacometrics \& Systems
  Pharmacology}} \textbf{\bibinfo{volume}{3}}, \bibinfo{pages}{148}
  (\bibinfo{year}{2014}).
\newblock
  \urlprefix\url{https://ascpt.onlinelibrary.wiley.com/doi/abs/10.1038/psp.2014.45}.

\bibitem{Jones2011}
\bibinfo{author}{Jones, H.~M.} \emph{et~al.}
\newblock \bibinfo{title}{Simulation of human intravenous and oral
  pharmacokinetics of 21 diverse compounds using physiologically based
  pharmacokinetic modelling.}
\newblock \emph{\bibinfo{journal}{Clinical pharmacokinetics}}
  \textbf{\bibinfo{volume}{50}}, \bibinfo{pages}{331--347}
  (\bibinfo{year}{2011}).

\bibitem{Wang2019}
\bibinfo{author}{Wang, Y.} \emph{et~al.}
\newblock \bibinfo{title}{Model-informed drug development: Current {US}
  regulatory practice and future considerations}.
\newblock \emph{\bibinfo{journal}{Clin. Pharmacol. Ther.}}
  \textbf{\bibinfo{volume}{105}}, \bibinfo{pages}{899--911}
  (\bibinfo{year}{2019}).

\bibitem{Mayr2016}
\bibinfo{author}{Mayr, A.}, \bibinfo{author}{Klambauer, G.},
  \bibinfo{author}{Unterthiner, T.} \& \bibinfo{author}{Hochreiter, S.}
\newblock \bibinfo{title}{Deeptox: Toxicity prediction using deep learning}.
\newblock \emph{\bibinfo{journal}{Frontiers in Environmental Science}}
  \textbf{\bibinfo{volume}{3}} (\bibinfo{year}{2016}).
\newblock
  \urlprefix\url{https://www.frontiersin.org/article/10.3389/fenvs.2015.00080}.

\bibitem{Fabian2020}
\bibinfo{author}{Fabian, B.} \emph{et~al.}
\newblock \bibinfo{title}{Molecular representation learning with language
  models and domain-relevant auxiliary tasks}.
\newblock \emph{\bibinfo{journal}{CoRR}}
  \textbf{\bibinfo{volume}{abs/2011.13230}} (\bibinfo{year}{2020}).
\newblock \urlprefix\url{https://arxiv.org/abs/2011.13230}.

\bibitem{GomezBombarelli2018}
\bibinfo{author}{Gómez-Bombarelli, R.} \emph{et~al.}
\newblock \bibinfo{title}{Automatic chemical design using a data-driven
  continuous representation of molecules}.
\newblock \emph{\bibinfo{journal}{ACS Central Science}}
  \textbf{\bibinfo{volume}{4}}, \bibinfo{pages}{268--276}
  (\bibinfo{year}{2018}).
\newblock \urlprefix\url{https://doi.org/10.1021/acscentsci.7b00572}.
\newblock \bibinfo{note}{PMID: 29532027}.

\bibitem{Winter2019}
\bibinfo{author}{Winter, R.}, \bibinfo{author}{Montanari, F.},
  \bibinfo{author}{Noé, F.} \& \bibinfo{author}{Clevert, D.-A.}
\newblock \bibinfo{title}{Learning continuous and data-driven molecular
  descriptors by translating equivalent chemical representations}.
\newblock \emph{\bibinfo{journal}{Chem. Sci.}} \textbf{\bibinfo{volume}{10}},
  \bibinfo{pages}{1692--1701} (\bibinfo{year}{2019}).
\newblock \urlprefix\url{http://dx.doi.org/10.1039/C8SC04175J}.

\bibitem{Winter2019a}
\bibinfo{author}{Winter, R.} \emph{et~al.}
\newblock \bibinfo{title}{Efficient multi-objective molecular optimization in a
  continuous latent space}.
\newblock \emph{\bibinfo{journal}{Chem. Sci.}} \textbf{\bibinfo{volume}{10}},
  \bibinfo{pages}{8016--8024} (\bibinfo{year}{2019}).
\newblock \urlprefix\url{http://dx.doi.org/10.1039/C9SC01928F}.

\bibitem{Obrezanova2022}
\bibinfo{author}{Obrezanova, O.} \emph{et~al.}
\newblock \bibinfo{title}{Prediction of in vivo pharmacokinetic parameters and
  time–exposure curves in rats using machine learning from the chemical
  structure} \textbf{\bibinfo{volume}{19}}, \bibinfo{pages}{1488--1504}
  (\bibinfo{year}{2022}).

\bibitem{Kingma2019}
\bibinfo{author}{Kingma, D.~P.} \& \bibinfo{author}{Welling, M.}
\newblock \bibinfo{title}{An introduction to variational autoencoders}
  (\bibinfo{year}{2019}).

\bibitem{Papamakarios2019}
\bibinfo{author}{Papamakarios, G.}, \bibinfo{author}{Nalisnick, E.},
  \bibinfo{author}{Rezende, D.~J.}, \bibinfo{author}{Mohamed, S.} \&
  \bibinfo{author}{Lakshminarayanan, B.}
\newblock \bibinfo{title}{Normalizing flows for probabilistic modeling and
  inference}.
\newblock \emph{\bibinfo{journal}{Journal of Machine Learning Research,
  22(57):1-64, 2021}}  (\bibinfo{year}{2019}).

\bibitem{Kobyzev2021}
\bibinfo{author}{Kobyzev, I.}, \bibinfo{author}{Prince, S. J.~D.} \&
  \bibinfo{author}{Brubaker, M.~A.}
\newblock \bibinfo{title}{Normalizing flows: An introduction and review of
  current methods} \textbf{\bibinfo{volume}{43}}, \bibinfo{pages}{3964--3979}
  (\bibinfo{year}{2021}).

\bibitem{Rackauckas2020}
\bibinfo{author}{Rackauckas, C.} \emph{et~al.}
\newblock \bibinfo{title}{Universal differential equations for scientific
  machine learning}  (\bibinfo{year}{2020}).

\bibitem{Raissi2017}
\bibinfo{author}{Raissi, M.}, \bibinfo{author}{Perdikaris, P.} \&
  \bibinfo{author}{Karniadakis, G.~E.}
\newblock \bibinfo{title}{Physics informed deep learning (part {I):}
  data-driven solutions of nonlinear partial differential equations}.
\newblock \emph{\bibinfo{journal}{CoRR}}
  \textbf{\bibinfo{volume}{abs/1711.10561}} (\bibinfo{year}{2017}).
\newblock \urlprefix\url{http://arxiv.org/abs/1711.10561}.

\bibitem{Raissi2017a}
\bibinfo{author}{Raissi, M.}, \bibinfo{author}{Perdikaris, P.} \&
  \bibinfo{author}{Karniadakis, G.~E.}
\newblock \bibinfo{title}{Physics informed deep learning (part {II):}
  data-driven discovery of nonlinear partial differential equations}.
\newblock \emph{\bibinfo{journal}{CoRR}}
  \textbf{\bibinfo{volume}{abs/1711.10566}} (\bibinfo{year}{2017}).
\newblock \urlprefix\url{http://arxiv.org/abs/1711.10566}.

\bibitem{Rackauckas2019}
\bibinfo{author}{Rackauckas, C.} \emph{et~al.}
\newblock \bibinfo{title}{Diffeqflux.jl-a julia library for neural differential
  equations}.
\newblock \emph{\bibinfo{journal}{arXiv preprint arXiv:1902.02376}}
  (\bibinfo{year}{2019}).

\bibitem{Merkelbach2022}
\bibinfo{author}{Merkelbach, K.} \emph{et~al.}
\newblock \bibinfo{title}{Hybridml: Open source platform for hybrid modeling}.
\newblock \emph{\bibinfo{journal}{Comput. Chem. Eng.}}
  \textbf{\bibinfo{volume}{160}}, \bibinfo{pages}{107736}
  (\bibinfo{year}{2022}).

\bibitem{Kidger2021}
\bibinfo{author}{Kidger, P.}
\newblock \emph{\bibinfo{title}{{O}n {N}eural {D}ifferential {E}quations}}.
\newblock Ph.D. thesis, \bibinfo{school}{University of Oxford}
  (\bibinfo{year}{2021}).

\bibitem{Ashukha2020}
\bibinfo{author}{Ashukha, A.}, \bibinfo{author}{Lyzhov, A.},
  \bibinfo{author}{Molchanov, D.} \& \bibinfo{author}{Vetrov, D.}
\newblock \bibinfo{title}{Pitfalls of in-domain uncertainty estimation and
  ensembling in deep learning}.
\newblock \emph{\bibinfo{journal}{arXiv e-prints}}
  \bibinfo{pages}{arXiv:2002.06470} (\bibinfo{year}{2020}).
\newblock
  \urlprefix\url{https://ui.adsabs.harvard.edu/abs/2020arXiv200206470A}.

\bibitem{ValentinJospin2020}
\bibinfo{author}{Valentin~Jospin, L.}, \bibinfo{author}{Buntine, W.},
  \bibinfo{author}{Boussaid, F.}, \bibinfo{author}{Laga, H.} \&
  \bibinfo{author}{Bennamoun, M.}
\newblock \bibinfo{title}{Hands-on bayesian neural networks -- a tutorial for
  deep learning users}.
\newblock \emph{\bibinfo{journal}{arXiv e-prints}}
  \bibinfo{pages}{arXiv:2007.06823} (\bibinfo{year}{2020}).
\newblock
  \urlprefix\url{https://ui.adsabs.harvard.edu/abs/2020arXiv200706823V}.

\bibitem{Gawlikowski2021}
\bibinfo{author}{Gawlikowski, J.} \emph{et~al.}
\newblock \bibinfo{title}{A survey of uncertainty in deep neural networks}.
\newblock \emph{\bibinfo{journal}{CoRR}}
  \textbf{\bibinfo{volume}{abs/2107.03342}} (\bibinfo{year}{2021}).
\newblock \urlprefix\url{https://arxiv.org/abs/2107.03342}.

\bibitem{Gruber2024}
\bibinfo{author}{Gruber, A.} \emph{et~al.}
\newblock \bibinfo{title}{Prediction of human pharmacokinetics from chemical
  structure: Combining mechanistic modeling with machine learning}.
\newblock \emph{\bibinfo{journal}{Journal of Pharmaceutical Sciences}}
  \textbf{\bibinfo{volume}{113}}, \bibinfo{pages}{55--63}
  (\bibinfo{year}{2024}).
\newblock
  \urlprefix\url{https://www.sciencedirect.com/science/article/pii/S0022354923004665}.

\bibitem{Peters2012}
\bibinfo{author}{Peters, S.~A.}
\newblock \bibinfo{title}{Physiologically-based pharmacokinetic (pbpk) modeling
  and simulations} (\bibinfo{year}{2012}).
\newblock
  \urlprefix\url{http://search.ebscohost.com/login.aspx?direct=true&scope=site&db=nlebk&db=nlabk&AN=448024}.
\newblock \bibinfo{note}{Description based upon print version of record}.

\end{thebibliography}

\end{document}